\documentstyle[12pt,psfig,aasms4]{article}
\received{}
\revised{}

\lefthead{Falceta-Gon\c{c}alves, D. and Jatenco-Pereira, V.}
\righthead{Acceleration of Dust Winds in Late-Type Stars}

\begin{document}

\title{THE EFFECTS OF ALFV\'EN WAVES AND RADIATION PRESSURE IN DUST WINDS OF 
LATE-TYPE STARS}

\author{Falceta-Gon\c{c}alves, D. and Jatenco-Pereira, V.}

\affil{Instituto de  Astronomia, Geof\'{\i}sica e Ci\^encias Atmosf\'ericas\\
 Universidade de S\~{a}o Paulo \\
Rua do Mat\~ao, 1226, 05508-900  S\~{a}o Paulo, SP, BRAZIL \\
diego@astro.iag.usp.br, jatenco@astro.iag.usp.br}

\begin{abstract}

In the present study, we analyze the effects of a flux of Alfv\'en waves 
acting together with radiation pressure on grains as an acceleration 
mechanism of the wind of late-type stars. In the wind model we simulate the 
presence of grains through a strong damping of the waves, we use a 
non-isothermal profile for temperature coherent with grain formation 
theories. We examine the changes in the velocity profile of the wind and 
we show that if the grains are created in the region 
$1.1 < r/r_0 < 2.0$ their presence
will affect the mass loss and terminal velocity. The model is 
applied to a K5 supergiant star and for Betelgeuse ($\alpha \, Ori$).


\end{abstract}

\clearpage

\section{Introduction}

Mass loss from late-type stars is evidenced by the blueshifted circumstellar
absorption lines present in their spectra. Giant and supergiant stars eject 
mass at a rate of $\sim 10^{-10} - 10^{-5} \; M_{\odot} \, yr^{-1}$  with 
terminal velocities $u_{\infty}  \sim 10 - 50 \; km \, s^{-1}$, which is lower 
than the surface escape velocity ($v_{e0}$) (\cite{deutsch56}; 
\cite{weymann62}; \cite{habing96}; \cite{lamerscassinelli99}; 
\cite{carpenter99}). An outward-directed flux of Alfv\'en waves has been 
proposed for driving these winds (\cite{belcherolbert75}; \cite{haisch80}; 
\cite{hartmannmacgregor80}; \cite{jpo89a}; \cite{macgregorcharbonneau94}; 
\cite{charbonneau95}). For late-type giant and supergiant stars, 
this nonthermal energy flux must be deposited in the region of subsonic 
flow to explain the observed $u_\infty \; < \; v_{e0}$ 
(\cite{hartmannmacgregor80}; \cite{leerholzer80}).

We have direct evidence for Alfv\'en waves in the solar wind.
Waves are observed with large perturbations in the magnetic
field, with negligible density perturbations, indicating the
presence of Alfv\'en waves. Since the early observation of MHD waves in the 
solar wind (\cite{belcherdavis71}), various authors have suggested that 
Alfv\'en waves could be important in transfer momentum to the wind.
We know from the solar 
coronal holes observations that the density is likely to 
fall off faster than $r^{-2}$ (e.g. \cite{hollweg81}; 
\cite{doschek97}), while the magnetic field strengths is 
proportional to $r^{-2}$ (\cite{linsky87}). In our 
simulations we show that this behavior exists up to 
$r < 3r_0$. The energy flux per 
unit area transported by the wave, $\phi_A$, is 
$\phi_A \simeq \epsilon v_A \simeq (1/2)\rho_{0}{\langle \delta v^2 \rangle} 
v_A$, where $\epsilon$ is the energy density of the waves. This energy 
flux is constant, 
when there is no damping, or decreases due to damping, such that 
$\epsilon$ decreases with $r$. Since the pressure associated with the wave 
is proportional to $\epsilon$, then this pressure also decreases with $r$. 
The result of this is a pressure gradient that accelerates the gas 
(\cite{castor81}). 

Alfv\'en waves have been used to explain the heating of 
stellar corona and the production of stellar winds in many 
regions of the Hertzsprung-Russell (HR) diagram. They are used to explain 
the solar wind acceleration (\cite{jpo89b}; \cite{jpoy94}), cool supergiant 
stars (\cite{charbonneau95}; \cite{airapetian00}) and winds in hot 
stars (\cite{cass79}; \cite{underhill83}; \cite{sjpo93}). These waves can 
also be important in T Tauri variables (\cite{hartmann82}; \cite{jpo89c}; 
\cite{vjpo00}). The origin of Alfv\'en waves was suggested by \cite{op86} 
to be the annihilation of twisted magnetic fields near the surface. These
annihilation sites that could also explain the variable hot-spots observed 
in late-type stars (Gray 2000, 2001; \cite{young00}). The waves can also
be generated by convective motions inner in the envelope, that create 
perturbations in the magnetic fields.

On the other hand, in cool supergiant stars, the presence of grains are well
know determined by infrared excesses (\cite{woolf69}). However, the region of
nucleation is not precisely determined. These grains are supposed to be the 
main responsible for radiation absorption, and also are supposed to contribute 
in the acceleration of the wind (e.g. \cite{jura86}; \cite{elitzur01}).
The temperature profile of the wind for these stars is also not very well 
known, but by nucleation theories (\cite{gail87}) the temperature range for 
grain formation 
is $400 \, K < T < 2500 \, K$. 

Besides, the high luminosity of these stars have lead many authors to 
include the radiation pressure as a mechanism to explain the mass loss. 
MacGregor and Stencel (1992) studied the interaction between gas and dust 
in the curcumstellar envelope of late-type stars. In their analysis the 
grains and gas are well coupled if the grain radius is $\sim 10^{-5} \; cm$ 
while for small grains $< 5 \times 10^{-6} \; cm$, the collision momentum 
transfer is insufficient to induce expansion of the atmosphere as a whole.

Jatenco-Pereira and Opher (1989a) (hereafter JPO), presented a model for mass
loss in a cool supergiant K5 star with $M = 16 M_{\odot}$, where a 
flux of Alfv\'en waves flux ($\phi_{A}$) 
is the main mechanism for accelerating the wind. The model solves the 
equations along the magnetic field axis, and assumes 
(i) a divergent geometry for magnetic field lines, 
(ii) a pure plasma (without dust) and (iii) isothermal wind. Holzer, Fla and Leer 
(1983) argued that wind driven by Alfv\'en waves from late-type stars 
only would be possible in an average ion-friction damping length 
$L_f \sim 0.85 - 1.0 r_0$. Since $L_f$ is proportional to $P^2$
(where $P$ is the average Alfv\'en period), this requires that 
nature fine-tune $P \sim 1.77 \times 10^4 \; s$. For long wave period 
($P > 10^4 \; s$), the wave damping length is large and the ion-neutral 
friction damping is not important (\cite{depontieu01}).
On the other hand, Davila (1985) emphasized that on the surface of the Sun 
one might expect that boundary effects may be important, and a WKB 
approximation may not be valid whenever the MHD wavelength ($\lambda$) 
becomes greater than $d \sim 10^{10} \; cm \; \sim 0.14 r_0$, 
the characteristic transverse dimension of a solar coronal hole, 
in despite of Usmanov et al. (2000) that used a WKB approximation for
the solar wind, in comparison with Ulysses data.
We assume an Alfv\'en wave flux with an averaged wavelength 
$\lambda < 0.5 r_0$ such that the WKB approximation is valid. 
JPO used a wave period $P > 10^4 \; s$ and 
discussed other possible damping mechanism for Alfv\'en waves such as 
nonlinear damping, resonant surface damping and turbulent damping. They 
show that these damping processes can produce acceptable winds in late-type 
giant stars. We extend the JPO model including a simulation of a strong 
damping of Alfv\'en waves, possibly due to a grain interaction, and the 
effect of radiation pressure in the wind acceleration. The changes in the 
velocity profile of the wind are compared with the pure plasma and isothermal 
model.

In section 2, we briefly describe the JPO model. In section 3, we present 
the inclusion of grain interaction with Alfv\'{e}n waves in a non-isothermal 
temperature profile and the radiation pressure on grains. In section 4, 
we apply this magnetic-radiation model in a K5 supergiant star and $\alpha \, 
Ori$ and discuss the results. In section 5 we summarize our conclusions.

\section{The Model}

The model suggested by JPO for mass loss in late-type stars takes into 
account our knowledge of coronal holes in the Sun. From surface, 
the magnetic field configuration is divergent, and after a 
transition radius, $r_T$, it becomes radial. The net non-radial expansion 
at this region is given by the factor of 
$F \equiv {\Omega \over \Omega_0} = 7.26   \; \; \;  (r < 3 R_\odot)$ 
(\cite{munrojackson77}). JPO used $F = 10$ for supergiant stars. They used 
the diverging geometry for the magnetic field suggested by 
\cite{kuinhearn82} and \cite{parker63} given by
\begin{equation}
\medskip
A(r) = A(r_0) \biggl( {r \over r_0} \biggr)^S \; ,
\medskip
\end{equation}
\noindent
where $A(r)$ is the cross section of the geometry at a radial
distance $r$, $r_0$ is the initial radius and $S$ is a parameter that 
determines the divergence of the geometry up to the transition radius $r_T$. 
The geometry is radial for $r > r_T$. JPO analyzed the effect of the change 
in the opening angle of the magnetic field on mass loss and terminal 
velocity of a \lq\lq typical" K5 supergiant star. They used a flux of 
Alfv\'en waves as the acceleration mechanism of the wind, with non-linear, 
surface Alfv\'en wave absorption and turbulent damping. In the model, for a 
given initial damping length at $r_0$ an upper limit of $S$ exist when mass 
loss stops, $S_{max}$. They found acceptable solutions for $3 < S < 7$. 
They showed for a range of initial damping lengths, $L_0 \sim 0.1 - 0.2 r_0$, 
and an isothermal atmosphere, $T \sim 10^4 \; K$, that a divergent geometry 
can produce the observed large mass loss rates and the small ratios 
$u_{\infty}/v_{e0} \sim 1/2$ of the supergiants.

The momentum equation of the wind may be deduced from continuity equation, 
conservation of momentum and action. The equation of motion (e.g., eq. [52] 
of \cite{Holzeretal83}; eq. [20] of JPO) is
\begin{equation}
{1 \over u} {du \over dr} = {Z \over r} {N
\over D} \, ,
\end{equation} 
\par\noindent
where $u$ is the ejected velocity at the radius $r$, $Z = S$ for $r_0 \leq r
\leq r_T$ and $Z = 2$ for $r > r_T$. $N$ and $D$ are given by
\begin{equation}
N \equiv v_{th}^2 \left (1 - {r \over v_{th}} {dv_{th} \over dr}
\right ) + {1 \over 4} \left ({{1 + 3M_A} \over {1 + M_A}} \right
) \langle \delta v^2 \rangle + {1 \over 4} {r \over L} \langle
\delta v^2 \rangle - {1 \over {2Z}} v_e^2 \, ,
\end{equation}
\begin{equation}
D \equiv u^2 - v_{th}^2 - {1 \over 4} \left ({1 +3M_A} \over {1 +
M_A} \right ) \langle \delta v^2 \rangle \, ,
\end{equation}
\par\noindent
where $v_{A}= B/\sqrt{4\pi \rho}$, is the Alfv\'{e}n velocity, 
$v_{th} = \sqrt{kT/(\mu m_{H})}$, is the thermal velocity, 
$M_{A} = u/v_{A}$, is the Mach number, $u$ is the wind velocity, 
$L$ is the damping length of wave and 
$\left\langle \delta v^{2}\right\rangle$ the velocity
associated to the magnetic field variation.

The equations were solved in 1-D, along the magnetic 
field lines, in a cylindrically symmetric geometry, assuming the wave 
flux consistent with $\delta B/B << 1$ and $P > 10^4 \; s$, 
which is implicit in the initial damping length. In the
present study the fluctuations in velocity, magnetic field and 
density are small allowing a linear treatment of Alfv\'en waves
(\cite{boynton96}), in despite of e.g. Ofman \& Davila (1998) model 
for solar coronal holes, where the fluctuations of density and 
velocity were large and a nonlinear treatment of the Alfv\'en waves 
were necessary.

\section{Grains simulation, temperature profile and radiation pressure}

\subsection{Simulation of strong damping of Alfv\'en waves}

In JPO model the Alfv\'en waves were supposed to be damped by nonlinear 
process with damping length, $L_1$, resonant surface damping, $L_2$, 
and turbulent damping, $L_3$, separately although they may contribute 
jointly. The damping length for each case is given by:

\begin{equation}
L_{1}=L_{10}\left[ \frac{\left\langle \delta v^{2}\right\rangle _{0}
}{\left\langle \delta v^{2}\right\rangle }\left( \frac{v_{A}}{v_{A0}}\right)
^{4}\left( 1+M_{A}\right) \right] \, ,
\end{equation}

\begin{equation} 
L_{2}=L_{20}\left( \frac{r_{0}}{r}\right) ^{\frac{S}{2}}\left( \frac{
v_{A}}{v_{A0}}\right) ^{2}\left( 1+M_{A}\right)\, ,
\end{equation}

\begin{equation}
L_{3}=L_{30}\left( \frac{u+v_{A}}{u_{0}+v_{A0}}\right) \left( \frac{r
}{r_{0}}\right) ^{\frac{S}{2}}\left( \frac{\left\langle \delta
v^{2}\right\rangle _{0}}{\left\langle \delta v^{2}\right\rangle }\right)^{1/2} 
\, ,
\end{equation} 
\par\noindent where $L_{10}$, $L_{20}$, $L_{30}$ are the initial damping 
length for each mechanism and $u_0$ is the initial plasma velocity.

The above damping lengths were used in pure hydrogen plasma, $i.e.$ there is
no grain formation. Experimental (\cite{elfimov97}), and theoretical 
(\cite{havnes89}), researches show that Alfv\'{e}n waves are suddenly damped 
in ``dust plasma'' environments. Impurities in the wind may increase the 
dissipation of the Alfv\'en waves. In the case of a stellar wind, 
inhomogeneous turbulent regions can occur immediately after the sonic 
point and they may manifest themselves as a sudden decrease in the damping 
length. We suppose that a sudden decrease in the damping length is 
correlated to the presence of grains although we do not treat them 
physically. This local dissipation may be simulated as an exponential 
decay of the damping length of the wave with distance given by:

\begin{equation}
\medskip
L_{grain} \equiv L_{r} = L_{r_{1}}\exp \frac{r_{1}-r}{A},
\medskip
\end{equation}
\noindent
where $r_{1}$ is the initial location in the wind where the grains can form 
which corresponds to the damping length $L_{r_{1}}$ and $A$ is a constant 
that shows how suddenly is the damping decay.

The determination of the grains formation region is highly uncertain, 
since this region can not be determined by the observations. 
Gail, Keller and Sedlmayr (1984), Gail and Sedlmayr (1986, 1987) and 
Clayton et al. (1992, 1995) 
presented a theoretical nucleation model. Their results determines the 
region $1r_{0} < r < 2r_{0}$ for SiO grains formation at cool supergiant 
envelopes, that is named here by ``grain formation point'' (GFP, hereafter).

Analyzing the dependence of the three previous damping mechanisms in 
the JPO model vs. $r$ (eqs. 5, 6 and 7) we note that all of them have the 
same behavior in the inner regions of the wind. They initially rise and 
peak for $r < 2r_{0}$ as we can see in Figure \ref{fig1}. It means that 
considering the region of nucleation given by Gail, $et$ $al.$ (1984), 
the model is independent of which initial damping mechanism we use, 
since differences only occur at large distances. For comparison 
we plot in Figure 1 these three damping lengths as a function of $r$ and the 
simulation of strong damping of Alfv\'en waves given by eq. 8. 
In our simulations we used the nonlinear mechanism ($L_1$) from the 
stellar surface until the region of grain formation and thereafter the 
wave damping assumes the form given in eq. 8.

\medskip
\centerline{[EDITOR PLACE FIGURE 1 HERE]}
\medskip

\subsection{The temperature profile}

Observations of the solar atmosphere show that the temperature 
($\sim 6000 \; K$) decreases reaching a minimum temperature at about 
$500 \; km$ above surface and increases 
slightly after, to $\sim 10^4 \; K$, into the chromosphere, and then 
suddenly jumps through a narrow transition layer to $\sim 2 \times 10^6 \, K$ 
in the corona. The solar wind is an extension
and a super-sonic expansion of the solar corona into interplanetary space 
(\cite{priest85}).

As in the Sun, we assume that the same behavior of temperature profile 
takes place in the giant and supergiant stars. This profile must be 
consistent with grain formation, since the grains can be formed only 
at low temperatures regions.

In this work we introduce a temperature profile from a minimum 
$\sim 2000 \; K$, above the photosphere, similar to the profile 
obtained by Hartmann \& Avrett (1984) and a rapidly increase to 
$\sim 10^{4} \; K$ at $r = 2r_{0}$. This profile is in agreement with 
the Gail and Sedlmayr (1986, 1987) theory. Rodgers and Glasshold (1991) 
modelled the thermal equilibrium of the alpha Ori atmosphere, 
assuming that heating occurs by grain-gas collisions and by 
radiation absorption and the cooling
occurs by adiabatic expansion and radiative de-excitation. They found a 
decreasing profile for $r > 2r_{0}$, 
from $10^{4}$ to $2000 \; K$ at $r = 10r_{0}$. By this way, we combined the 
solar observations, for sudden increase of temperature in chromosphere, 
to the Rodgers \& Glasshold (1991) model.

\subsection{Radiation pressure}

Many authors have tried to explain the mass loss processes of luminous 
stars like Wolf-Rayet using only radiation pressure as the accelaration 
mechanism. The great problem that arises in this star
is called ``the momentum problem'', that consists of a radiation field 
momentum lower than the momentum of the wind 
(\cite{abbott78}): $\eta = \dot{M}u_{\infty }/( L_{*}/{c})$ is in the 
range $3 - 30$ (\cite{barlowsmithwillis81}). 
dos Santos, Jatenco-Pereira \& Opher (1993) presented a model for mass 
loss in Wolf-Rayet stars where a flux of Alfv\'en waves acting together 
with radiation pressure has enough momentum flux and energy flux to 
drive the wind. They assumed a plasma without dust and isothermal wind.

In order to verify if radiation pressure can be important in driving 
late-type star wind we include the radiation pressure into the wind 
equation using an effective escape velocity.
Taking a pure radiative driven wind, the force 
exerted on a grain of radius $a$ situated at a distance 
$r$ from a star of luminosity $L_\star$ and mass 
$M_\star$ is:

\begin{equation}
F = \frac{\pi a^2 Q L} {4 \pi r^2 c} \; ,
\end{equation}

\par\noindent
where $Q$ is the mean extinction efficiency factor of the grain, and 
$a \sim 10^{-5} \; cm$ is the grain radius value.

	Using the Newton's Law, the wind equation becomes:

\begin{equation}
u \frac{du} {dr} + \frac{1} {\rho} \frac{dP} {dr} = - \frac{GM_\star} {r^2} +
\frac{n_d} {\rho} F \; ,
\end{equation}

\par\noindent
or

\begin{equation}
u \frac{du} {dr} + \frac{1} {\rho} \frac{dP} {dr} + \frac{GM_\star} {r^2} 
(1 - \Gamma) = 0 \; ,
\end{equation}

\par\noindent 
where the factor $\Gamma$ is given by:
$\Gamma =\frac{n_d} {\rho} \frac{\pi a^2 Q L_\star} {4 \pi c G M_\star}$, 
with $\frac{n_d} {\rho}$ the ratio of the grain number density 
to the gas mass density.

If we call:

\begin{equation}
- \frac{1} {2}(v_{esc})^2 = -\frac{GM_\star} {r} (1 - \Gamma) \; ,
\end{equation}

\par\noindent 
we define the effective escape velocity..

The escape velocity (12) is introduced in equation (3) in the 
region where the grains may affect the wind 
dynamics i.e. for $r > GFP$ (with $1.1 r_0 < GFP < 2r_0$).
We assume that most of the radiation is absorbed by the grains 
in this region. This radiation absorption 
is partially responsible for grain destruction and acceleration. 
These accelerated grains are dynamically connected to gas 
(\cite{lafon91}; \cite{liberatore01}), so that the grains transfer momentum 
to the gas inducing the mass loss.

\section{Results}

This magnetic-radiation driven winds model consider a steady, cylindrically 
symmetric and 1-D wind emanating 
from a divergent magnetic field geometry with a temperature gradient 
similar to the Sun atmosphere. We assume that a flux of Alfv\'en waves, which 
properties were defined in earlier sections, are generated near the stellar 
surface and propagate outward. Since in 
the inner region of the wind the damping mechanisms already studied in 
JPO have the same behavior we so use the non-linear one untill the GFP, after 
that we use the damping length given by eq. 8. In the region of grain 
simulation ($r > GFP$) the radiation 
pressure in grains, as well as the grain damping, are taken into account
from GFP to $300 r_0$. We applied the 
model to the K5 supergiant, (model 6 of Hartmann \& MacGregor (1980)) 
and $\alpha \, Ori$.

\subsection{K5 supergiant star}

We used $M_{*}=16 M_{\odot }$, $B=10 \; G$, $r_{0}=400 R_{\odot}$, 
$L_{*}=3 \times 10^{4} L_{\odot}$, $\phi_{A0}=3.36 \times 10^{6} \; erg \; 
cm^{-2} \; s^{-1}$, $\Gamma =0.3$, $S = 5$, $A = 0.1$, 
the initial damping length $L_0 = 0.2r_0$ and the density on wind basis 
$\rho =  10^{-13} g cm^{-3}$. The typical velocity 
and mass loss rate observed for these stars are of 
order $u_{\infty} \sim \frac{1} {2} v_{e0}$ and 
$\dot{M} = 5 \times 10^{-7} \; M_{\odot} \; yr^{-1}$ 
(see e.g. \cite{charbonneau95}; \cite{whitelock95}; 
\cite{lamerscassinelli99}; \cite{lindqvist00}).

First we performed some simulations in an isothermal wind in order to 
evaluate the influence of the grains on the terminal 
velocities, found by previous pure plasma model. The 
introduction of damping by grains has an important effect in decreasing the 
flux of Alfv\'en waves much more effectively than the others damping 
mechanisms. As a consequence the terminal velocity of the wind reaches lower 
values as compared with JPO model. Since the grain formation region is not 
well known, we introduced the damping by grains at different distances 
from the stellar surface. The models show that as closer to the surface the
grain if formed, lower is the asymptotic velocity. This effect is 
caused by the lose of wave energy by the strong damping, which leaves the wind 
unsupported against gravity. Decreasing the non-radial expansion factor 
$S$ we can collimate the wave flux into a smaller area accelerating the gas 
more efficiently. On the other hand, we found that for $r > 2r_0$ the 
presence of grains is not important for the wind dynamics. The properties 
of the model can be seen in Table 1.

\medskip
\centerline{[EDITOR PLACE TABLE 1 HERE]}
\medskip

The velocity curve obtained by solving the equation of the wind (2) is plotted 
in Figure \ref{fig2}. For comparison we plot the velocity profile
for a pure plasma and isothermal 
wind (dotted line), and for the magnetic-radiation model 
(solid line). We may note that the terminal
velocity for the magnetic-radiation model is higher than the JPO concluding 
then that the 
introduction of grain and temperature profile affect the wind dynamics. The 
grain presence diminishes the asymptotic velocity, while the introduction 
of a temperature gradient increases the velocity. For a given density, 
temperature, magnetic field and GFP (which are all observable) we may 
adjust the model to the expected values of $u/v_{eo}$ and $\dot M$, by 
varying the Alfv\'en wave flux. The best fit occurs for an initial Alfv\'en 
wave flux of $5 \times 10^6 \; erg \; cm^{-2} \; s^{-1}$. On the figure 2, 
for a fixed GFP, we note that the terminal velocity is higher for lower S 
values due to the consideration of a radiative and non-isothermal
atmosphere. In this sense increasing the S value we should have 
a decrease in the terminal velocity. However, for higher 
S values there was no sonic point. The observations can be 
reproduced if we increase the initial Alfv\'en flux to 
$5 \times 10^6 \; erg \; cm^{-2} \; s^{-1}$. 
In the figure we mark the region where 
we assume the formation of grains.

\medskip
\centerline{[EDITOR PLACE FIGURE 2 HERE]}
\medskip

Although for supergiant stars the momentum problem is not present, it is 
interesting to estimate the value of $\eta$. For example, for K5 supergiant 
star using the above values we have $\eta \sim 0.4$, lower than 1, but still 
high considering normal absorption coefficients. Using the above values for 
the K5 supergiant star, we can estimate the ratio ($\xi$) of the initial 
energy flux of Alfv\'{e}n waves to the radiation energy flux: 
$\xi = \phi_{A0}/(L_{*}/(4\pi r_{0}^{2})) \sim 10^{-3}$. This value 
indicates that the energy flux of Alfv\'{e}n waves is only a few percent of 
the stellar luminosity. This relation may be used to estimate the Alfv\'{e}n 
wave flux on other stars like Betelgeuse, for example.

\subsection{Betelgeuse ($\alpha \, Ori$)}

Betelgeuse observations give the terminal velocity, $u_{\infty}  \sim 15 \; 
km \; s^{-1}$ and a mass loss rate $\dot{M} \sim 2 - 5 \times 10^{-6} \; 
M_{\odot } \; yr^{-1}$ (\cite{loup93}; \cite{knapp91}). We applied the 
magnetic-radiation model to this star with a temperature profile given by 
Rodgers and Glasshold (1991). We used the following initial conditions: 
$M_{*} = 8 M_{\odot}$, $B = 10 \; G$, $r_{0} = 300 R_{\odot}$, 
$L_{*} = 10^{4} L_{\odot}$ and $\Gamma = 0.7$, $S = 5$, $L_{0} = 0.2r_{0}$ 
and the GFP located at $r = 1.24 r_0$. The Alfv\'{e}n wave flux can be 
determined by $\xi = 10^{-3}$. We found $\phi_{A0} = 6.9 \times 10^{6} \; 
erg \; cm^{-2} \; s^{-1}$. As in the previous section, we performed the 
calculations until $r = 300r_o$. However, we show in 
Figure 3 the resultant velocity profile until $r = 10r_0$
since for $r > 10r_0$ the curve becomes flattened showing 
the same terminal velocity.

\medskip
\centerline{[EDITOR PLACE FIGURE 3 HERE]}
\medskip

We obtained $u_{\infty} \sim 16 \; km \; s^{-1}$ and $\dot{M} = 4 \times 
10^{-6} \; M_{\odot} \; yr^{-1}$, very similar to the observations. 
Betelgeuse also has high UV emission at $r \sim 2r_{0}$ region. 
This emission may be connected to the MHD waves damping (\cite{gilliand96}) 
and with a sudden heating of this region, which is in agreement with our model.

\section{Conclusions}

In this work we present a model to study mass loss processes in
late-type stars, using an Alfv\'{e}n waves flux added to radiation pressure as
accelerating mechanisms. We simulate the presence of grains in the sense of 
a strong damping of Alfv\'en waves in the region 
$r > GFP$ (which is about $1.1r_0 < GFP < 2r_0$).
In this region we added the radiation pressure as an acceleration 
mechanism of the wind. In accordance with grain formation theories and in 
agreement with observations, we took into account a temperature profile for 
the wind going from $\sim 2000 \; K$ above the photosphere,
with a rapidly grown to $\sim 10^{4} K$ at $r = 2r_{0}$ 
and a decreasing profile for $r > 2r_0$. The models show that
as closer as to the surface the grain is formed, lower is the
asymptotic velocity and that 
the wind velocity is not changed if the region of grain formation is located 
above $2r_0$. Although a number of hypotheses have been made this 
hybrid model can lead to a more realistic wind and gives more constraint 
in the physical parameters of the wind. The model was applied to a 
K5 supergiant star and Betelgeuse, resulting in values that agrees with 
observations. Further studies will include the physical parameters of the 
grains as well as their distribution in the wind.

\smallskip

\acknowledgements
D.F.G and V.J.P. would like to thank the Brazilian agency CNPq for 
financial and partial support, respectively. D.F.G. would like to thank 
the Brazilian agency FAPESP for providing computational equipment 
under grant (No. 96/00677-3). The authors would like also to thank the 
project PRONEX/FINEP (No. 41.96.0908.00) for partial support.

\newpage

\begin{center}
{\bf Figure captions}
\end{center}

\figcaption[f1.eps]{The behavior of damping length as a function of the 
distance for resonant damping (solid curve), nonlinear damping 
(dashed curve), turbulent damping (dashed-dot curve) and by grains 
(dotted curve). The initial damping length is $L_0 = 0.2r_0$ for the 
three cases. For damping by grains (eq. 8) we used $r_1 = 1.32r_0$, 
$L_{r1} = 0.75r_0$ and $A = 0.1$.
\label{fig1}}

\figcaption[f2.eps]{Velocity profile for JPO homogeneous and isothermal 
model (dotted line) and for the hybrid model including grain presence, 
non-isothermal atmosphere and radiation pressure (solid line). The grain 
formation region is marked in the figure. \label{fig2}}

\figcaption[f3.eps]{Velocity profile as a function of distance for 
$\alpha \, Ori$. In this non-isothermal model we used a flux of Alfv\'{e}n 
waves added to radiation pressure as an acceleration mechanism for the wind. 
\label{fig3}}


\newpage

\begin{thebibliography}{}
 
\bibitem[Abbott 1978]{abbott78} Abbott, D.C. 1978, \apj, 225, 893

\bibitem[Airapetian et al. 2000]{airapetian00} Airapetian, V. S.;
Ofman, L.; robinson, R. D.; Carpenter, K. \& Davila, J. 2000, \apj, 528, 965

\bibitem[Barlow, Smith \& Willis 1981]{barlowsmithwillis81} Barlow M. J., 
Smith, L. J. \& Willis, A. J. 1981, MNRAS, 196, 101

\bibitem[Belcher \& Davis 1971]{belcherdavis71} Belcher,J. W. \& 
Davis, L. Jr. 1971, J. Geophys. Res., 76, 3534

\bibitem[Belcher \& Olbert 1975]{belcherolbert75} Belcher,J. W. 
\& Olbert, S. 1975, \apj, 200, 369

\bibitem[Boynton \& Torkelsson 1996]{boynton96} Boynton, G. C.,
Torkelsson, U. 1996, A\&A, 308, 299

\bibitem[Clayton et al. 1992]{} Clayton, G. C., Whitney, B. A.,
Stanford, S. A., Drilling, J. S. 1992, \apj, 397, 652

\bibitem[Clayton et al. 1995]{} Clayton, G. C., Kelly, D. M.,
Lacy, J. H., Little-Maressin, I. R., Feldman, P. A., Bernath, P. F.
1995, AJ, 109, 2096

\bibitem[Carpenter et al. 1999]{carpenter99} Carpenter, K. G., Robinson,
R. D., Harper, G. M., Bennett, P. D., Brown, A.k Mullan, D. J. 1999, \apj,
521m 382

\bibitem[Cassinelli 1979]{cass79} Cassinelli, J. P. 1979, ARAA, 17, 275

\bibitem[Castor 1981]{castor81} Castor, J. J. 1981, \lq \lq Physical 
Processes in Red Giants", eds. I. Iben Jr., A. Renzini, Dordrecht, Reidel

\bibitem[Charbonneau \& Macgregor 1995]{charbonneau95} 
Charbonneau, P. \& Macgregor, K. B. 1995, \apj, 454, 901

\bibitem[Davila 1985]{davila85} Davila, J. M. 1985, \apj, 291, 328

\bibitem[Deutsch 1956]{deutsch56} Deutsch, A. J. 1956, \apj, 123, 210

\bibitem[De Pontieu, Martens \& Hudson 2001]{depontieu01} De Pontieu, B.,
Martens, P. \& Hudson, H. 2001, \apj, 558, 859

\bibitem[dos Santos, Jatenco-Pereira, \& Opher 1993]{sjpo93} 
dos Santos, L.C., Jatenco-Pereira, V., \& Opher, R. 1993, \apj, 81, 119

\bibitem[Doschek et al. 1997]{doschek97} Doschek, G. A. et al. 1997, \apj,
482, L109

\bibitem[Elfimov et al. 1997]
{elfimov97} Elfimov, A. G., Galv\~{a}o, R. O., Nascimento, I. C., 
\& Amarante-Segundo, G. 1997, Plasm. Phys. Control. Fusion, 39, 1551

\bibitem[Elitzur \& Ivezic 2001]{elitzur01} Elitzur, M. \& Ivezic, Z. 2001,
MNRAS, 327, 403

\bibitem[Gail, Keller \& Sedlmayr 1984]{gail84} Gail, H.P., Keller, R. 
\& Sedlmayr, E. 1984, \aap, 320, 332

\bibitem[Gail \& Sedlmayr 1986]{gail86} Gail, H.P. \& Sedlmayr, E. 1986, 
\aap, 148, 183

\bibitem[Gail \& Sedlmayr 1987]{gail87} Gail, H.P. \& Sedlmayr, E. 1987, 
\aap, 177, 186

\bibitem[Gilliand \& Dupree 1996]{gilliand96} Gilliand, R. \& Dupree, A. 
1996, \apj, 463, L29

\bibitem[Gray 2000]{gray00} Gray, D. 2000, \apj, 532, 487

\bibitem[Gray 2001]{gray01} Gray, D. 2001, PASP, 113, 1378

\bibitem[Habing 1996]{habing96} Habing, H. 1996, A\&AR, 7, 25

\bibitem[Haisch, Linsky \& Basri 1980]{haisch80} Haisch, B. M., Linsky, J. L. 
\& Basri, G. S. 1980, \apj, 235, 519

\bibitem[Hartmann \& Avrett 1984]{} Hartmann, L. \& Avrett, E. 1984, \apj,
284, 238

\bibitem[Hartmann, Edwards \& Avrett 1982]{hartmann82} Hartmann, L., Edwards, 
S. \& Avrett, E. 1982, \apj, 261, 279

\bibitem[Hartmann \& MacGregor 1980]{hartmannmacgregor80} Hartmann, L. 
\& MacGregor, K.B. 1980, \apj, 242, 260

\bibitem[Havnes, Hartquist \& Phillip 1989]{havnes89} Havnes, O., Hartquist, 
T. \& Phillip, W. 1989, \aap, 217, L13

\bibitem[Holzer, Fla \& Leer 1983]{Holzeretal83}Holzer, T. E., Fla, T. 
\& Leer, E. 1983 \apj, 275, 808

\bibitem[Hollweg 1981]{hollweg81} Hollweg, J. V. 1981, Solar Phys., 70, 25

\bibitem[Jatenco-Pereira \& Opher 1989a]{jpo89a} Jatenco-Pereira, V. 
\& Opher, R. 1989a, \aap, 209, 327 (JPO)

\bibitem[Jatenco-Pereira \& Opher 1989b]{jpo89b} Jatenco-Pereira, V. 
\& Opher, R. 1989b, \apj, 344, 513

\bibitem[Jatenco-Pereira \& Opher 1989c]{jpo89c} Jatenco-Pereira, V. 
\& Opher, R. 1989c, MNRAS, 236, 1

\bibitem[Jatenco-Pereira, Opher \& Yamamoto 1994]{jpoy94} Jatenco-Pereira, 
V., Opher, R. \& Yamamoto, L. C, 1994, \apj, 432, 409

\bibitem[Jura 1986]{jura86} Jura, M. 1986, \apj, 303, 327

\bibitem[Knapp \& Woodhams 1991]{knapp91} Knapp, G. \& Woodhams, M. 1991, 
ASP Conf. Series, 35, 199

\bibitem[Kuin \& Hearn (1982)]{kuinhearn82} Kuin, N. P. M. \& Hearn, A., G. 
1982, \aap, 114, 303

\bibitem[Lafon \& Berruyer 1991]{lafon91} Lafon, J., Berruyer, N. 1991,
A\&AR, 2, 249

\bibitem[Lamers \& Cassinelli 1999]{lamerscassinelli99} Lamers, H. J. G. L. M.
\& Cassinelli, J. P. 1999, \lq\lq Introduction to Stellar Winds'', H.J.G.L.M.
Lamers and J.P. Cassinelli (eds.), Cambridge University Press

\bibitem[Leer \& Holzer 1980]{leerholzer80} Leer, E. \& Holzer, T. E. 
1980, J. Geophys. Res., 85, 4681

\bibitem[Liberatore, Lafon \& Berruyer 2001]{liberatore01} Liberatore, S.,
Lafon, J. \& Berruyer, N. 2001, A\&A, 377, 522

\bibitem[Lindqvist et al. 2000]{lindqvist00} Lindqvist, M.; Shoier, F. L.; 
Lucas, R. \& Olofsson, H. 2000, A\&A, 361, 1036

\bibitem[Linsky 1987]{linsky87} Linsky, J. L. 1987, \lq\lq Circunstellar
Matter'', I. Appenzeller and C. Jordan (eds.)

\bibitem[Loup et al. 1993]{loup93} Loup, C., 
Forveille, T., Omont, A. \& Paul, J. 1993, \aap, 99, 291

\bibitem[MacGregor \& Charbonneau 1994]{macgregorcharbonneau94} 
MacGregor, K. B. \& Charbonneau, P. 1994, \apj, 430, 387

\bibitem[MacGregor \& Stencel 1992]{macgregorstencel92} MacGregor, K. B. 
\& Stencel R. E. 1992, \apj, 397, 644

\bibitem[Munro \& Jackson 1977]{munrojackson77} Munro, R. H. \& Jackson, B. 
V. 1977, \apj, 213, 874

\bibitem[Ofman \& Davila 1998]{ofmandavila98} Ofman \& Davila, J. M. 1998,
JGR, 103, 23677

\bibitem[Opher \& Pereira (1986)]{op86} Opher, R. \& Pereira, V. J. S. 
1986, \apj, L107

\bibitem[Parker (1963)]{parker63} Parker, E. N. 1963 \lq \lq Interplanetary 
Dynamical Processes", Wiley, New York

\bibitem[Priest 1985]{priest85} Priest, E. R. 1985 \lq \lq  Solar System 
Magnetic Fields", D. Reidel

\bibitem[Rodgers \& Glasshold 1991]{rodgers91} Rodgers, B. 
\& Glasshold, A.E. 1991, \apj, 382, 606

\bibitem[Underhill 1983]{underhill83} Underhill, A. B. 1983, \apj, 268, L127

\bibitem[Usmanov et al. 2000]{} Usmanov, A. V., Goldstein, M. L., 
Besser, B. P., Fritzer, J. M. 2000, JGR, 105, 12675

\bibitem[Vasconcelos, Jatenco-Pereira \& Opher 2000]{vjpo00} Vasconcelos, 
M. J., Jatenco-Pereira, V. \& Opher, R. 2000, \apj, 534, 967

\bibitem[Weymann 1962]{weymann62} Weymann, R. 1962, \apj, 136, 844

\bibitem[Whitelock et al. 1995]{whitelock95} Whitelock, P.; Menzies, J.; 
Feast, M., Catchpole, R.; Marang, F. \& Carter, B. 1995, Ap\&SS, 230, 495

\bibitem[Woolf \& Ney 1969]{woolf69} Woolf, N. J. \& Ney, E. P. 1969, 
\apj, 155, 181

\bibitem[Young, Baldwin, Boysen et al. 2000]{young00} Young, J., Baldwin, J.,
Boysen, R. et al. 2000, MNRAS, 315, 635

\end{thebibliography}
\end{document}